\documentclass[11pt]{article}
\usepackage{amssymb,epsf,latexsym, amsmath}
\newcommand{\bfi}{\bfseries\itshape}

\makeatletter
\@addtoreset{equation}{section}

\makeatother

\newtheorem{thm}{Theorem}[section]

\newtheorem{lemma}[thm]{Lemma}

\allowdisplaybreaks
\begin{document}

\title{A Nonlinear Analysis of the Averaged Euler Equations}

\author{
{\small Darryl D. Holm}
\\{\small T-Division and CNLS, MS-B284}
\\{\small Los Alamos National Laboratory}
\\{\small Los Alamos, NM 87545 USA}
\\{\footnotesize email: dholm@lanl.gov}
\and {\small Shinar Kouranbaeva}
\\{\small Department of Mathematics}
\\{\small University of California Santa Cruz}
\\{\small Santa Cruz, CA 92093 USA}
\\{\footnotesize email: shinar@math.ucsc.edu}
\and {\small Jerrold E. Marsden}
\\{\small Control and Dynamical Systems}
\\{\small Caltech 107-81}
\\{\small Pasadena, CA 91125 USA}
\\{\footnotesize email: marsden@cds.caltech.edu}
\and  {\small Tudor Ratiu}
\\{\small Departement de Mathematiques}
\\{\small Ecole Polytechnique federale de Lausanne}
\\{\small CH - 1015 Lausanne, Switzerland}
\\{\footnotesize email: ratiu@math.ucsc.edu}
\and {\small Steve Shkoller}
\\{\small Center for Nonlinear Studies, MS-B258}
\\{\small Los Alamos, NM 87545 USA}
\\{\footnotesize email: shkoller@cds.caltech.edu}
}
\date{May 1998; this version, February 24, 1999\\[12pt]
{\it Dedicated to Vladimir Arnold on the occasion of his
60th birthday}}
\maketitle

\begin{abstract}
This paper develops the geometry and analysis of the
averaged Euler equations for ideal incompressible flow in domains in
Euclidean space and on Riemannian manifolds, possibly with boundary.
The averaged Euler equations involve a parameter $\alpha$; one
interpretation is that they are obtained  by ensemble averaging the Euler
equations in Lagrangian representation over rapid fluctuations whose
amplitudes are of order $\alpha$. The particle flows associated with these
equations are shown to be geodesics on a suitable group of volume
preserving diffeomorphisms, just as with the Euler equations themselves
(according to Arnold's theorem), but with respect to a right invariant
$H^1$ metric instead of the $L^2$ metric. The equations are also equivalent
to those for a certain second grade fluid. Additional properties of the
Euler equations, such as smoothness of the geodesic spray (the Ebin-Marsden
theorem) are also shown to hold. Using this nonlinear analysis framework,
the limit of zero viscosity for the corresponding viscous equations is
shown to be a regular limit, {\it even in the presence of boundaries}.
\end{abstract}

\newpage
\tableofcontents

\section{Introduction}

More than twenty-five years ago, using a setting introduced by
Arnold [1966], Ebin and Marsden [1970] proved that on manifolds with
no boundary (such as spatially periodic flow in Euclidean three
space ${\mathbb R}^3$), the solutions of the Navier-Stokes equation
converge to those of the Euler equations as the viscosity tends to
zero. Marsden, Ebin, and Fischer [1972] conjectured that although
in a region with boundary, solutions of the Navier-Stokes equations
would {\it not} in general converge to the solutions of the Euler
equations, a certain averaged quantity of the flow may converge.
Kato [1984] showed that the problem in the context of weak solutions
of the Navier-Stokes equations has fundamental difficulties.

Recently, Barenblatt and Chorin [1998a,b] also speculated that
certain average properties of the flow possess well-defined limits as
the viscosity tends to zero. One of the main purposes of this
paper is to develop extensions of the tools of Arnold, Ebin and
Marsden and to use them to prove that, in a sense that will be made
precise, {\it viscous flow, with an appropriate ensemble averaging
over rapid fluctuations or spatial averaging over small scales, does
indeed converge to solutions of the corresponding inviscid limit equations,
even in the presence of boundaries.}

\paragraph{Some History.} The geometric approach to fluid
mechanics has a long and complex history, going back at least to
the basic work of Poincar\'{e} (see the bibliographical
references, but especially [1901b, 1910]) and the thesis of
Ehrenfest (see Klein [1970]). In more recent times, it was the
work of Arnold [1966] that was a critical contribution.  Arnold
showed, amongst other things, that if $u(x,t)$ is a time dependent
divergence free vector field on a compact Riemannian
$n$-manifold $M$, possibly with boundary, if $u$ is parallel to the
boundary and if $\eta(x,t)$ is its volume preserving flow, then $u$
satisfies the Euler equations
\[
\frac{\partial u}{\partial t}  + \nabla _u u = - \text{grad} \ p
\]
if and only if the curve $t \mapsto \eta (\cdot, t)$ is an $L^2$
geodesic in $\mathcal{D}_\mu(M)$, the group of $C^{\infty}$ volume
preserving diffeomorphisms of $M$. Of course, the Euler equations in
Euclidean space in coordinates are given as follows (using the
summation convention for repeated indices):
\[
\frac{\partial u^i}{\partial t}
+ u^j
\frac{\partial u^i }{\partial x^j}
= - \frac{\partial p}{\partial x^i}\, ,
\]
and on a Riemannian manifold, the Euler equations take the
following form in coordinates:
\[
\frac{\partial u^i}{\partial t}
+ u^j \frac{\partial u^i }{\partial x^j}
+ \Gamma^i_{jk} u ^j u ^k
= - g^{ij}\frac{\partial p}{\partial x ^j}\,,
\]
where $g _{ij}$ is the Riemannian structure, $g ^{ij}$ is the
inverse metric tensor and $\Gamma^i_{jk}$ are the associated
Christoffel symbols.

Ebin and Marsden [1970] proved a remarkable result: the geodesic
spray of the $L^2$ right invariant  metric on $\mathcal{D}_\mu^s(M)$,
the group of volume preserving Sobolev $H^s$ diffeomorphisms ($s >
(n/2) + 1$) is $C^{\infty}$ (recall that $n$ is the dimension of the
underlying manifold $M$). They derived a number of interesting consequences
from this result, including theorems on the convergence of solutions of the
Navier-Stokes equations to solutions of the Euler equations as the
viscosity goes to zero when $M$ has no boundary. Since that time, several
papers (which we shall not review here) reproved and in some cases,
rediscovered, these results using more traditional PDE methods in Eulerian
representation. However, the depth and beauty of the direct approach in
Lagrangian representation remains compelling.

There have been, of course, many other developments since that
time, such as developments in the theory of hydrodynamic
stability, bifurcations, and other aspects of the geometric
approach. We refer to Arnold and Khesin [1998], Holm, Marsden,
Ratiu and Weinstein [1985], Marsden [1992] and Marsden and Ratiu [1999] and
references therein for stability studies and to Misiolek [1993, 1996] and
Shkoller [1998] and references therein for the Riemannian
geometry developments of the group of diffeomorphisms.

\paragraph{$H^1$-Geodesics.} The ($\kappa = 0$) Camassa-Holm
equation in one spatial dimension is given by
\begin{equation} \label{1dch.eqn}
u_t-u_{xxt}=-3uu_x+2u_x u_{xx}+uu_{xxx},
\end{equation}
or equivalently,
\[
\frac{\partial v}{\partial t}  + u v_x + 2 v u_x = 0\,
\]
where $v = u - u_{xx}$. This equation has attracted much attention;
amongst many other features, it is a completely integrable
bi-Hamiltonian system with non-smooth solitons that have very
interesting geometry. See, for example,  Camassa and Holm [1993], Camassa,
Holm and Hyman [1994], and Alber, Camassa, Holm and Marsden [1994]. This
one dimensional equation was given an Arnold-type interpretation in
terms of $H^1$ geodesics on an infinite dimensional group by Misiolek
[1998] (who also considered the case $\kappa \neq 0$ and used the
Bott-Virasoro group) and Kouranbaeva [1999] (who explored the geometry of
the $\kappa = 0$ equations using the diffeomorphism group) and Shkoller
[1998] for the well posedness in $H ^{3/2 + \epsilon}( S^1)$ (See also
Theorem \ref{1dch.thm} of the present paper).

\paragraph{The Averaged Euler Equations.} Using recent
developments in Lagrangian reduction and the Euler--Poincar\'e
equations, Holm, Marsden and Ratiu [1998a,b,c],
developed the ``Arnold view'' of fluid mechanics further by
applying it to many other types of fluid equations, such as
those in geophysics. One of the equations they studied is quite
remarkable and is the subject of this work.

The equations of concern to us may be described
in two mathematically equivalent ways. This equivalence is the
$H^1$-analogue of the theorem of Arnold mentioned above:
\begin{enumerate}
\item  First of all, of course, there are the explicit
{\bfi averaged Euler equations} (these are written in Euclidean
coordinates on Euclidean $n$-space; we shall write the equations on
manifolds in the main text):
\begin{equation}
\label{mme.equations}
\frac{\partial v^i}{\partial t}
+ u^j \frac{\partial v^i}{\partial x^j}
-\alpha^2
\left[
\frac{\partial u^j}{\partial x^i}
\right]
\triangle u_j = - \frac{\partial p}{\partial x ^i}\,,
\end{equation}
where $\alpha$ is a positive
constant, $v = u - \alpha^2\triangle u$ and $\triangle$ denotes the
componentwise Laplacian, and there is a summation over repeated indices
(and in Euclidean coordinates, there is no difference between indices
up or down).

As with the usual Euler equations, the function
$p$ is determined from the condition of incompressibility:
$\operatorname{div}u = 0$; the pressure satisfies a Poisson equation
that is determined by taking the divergence of the equation.

The no slip boundary conditions $u = 0$ on the boundary will
be assumed for equation (\ref{mme.equations}). A second
possible choice will be briefly discussed later.
\item The second description is to say that the flow $\eta_t(
\cdot) : = \eta ( \cdot, t)$ of the time dependent vector field
$u$ is a geodesic in a subgroup of $\mathcal{D}_\mu^s(M)$ with
respect to the right invariant $H^1$ metric.
\end{enumerate}
There are two interesting and
distinct ways of deriving the equations from the point of view of
mechanics:
\begin{enumerate}
\item
The first derivation is by means of averaging the Euler
equations over rapid fluctuations whose amplitude is of order $\alpha$ (see
Holm, Marsden and Ratiu [1998b] for the case of Euclidean space). This
method is explored in Marsden, Ratiu and Shkoller [1999b].
\item
The second derivation uses the notion of second grade fluids, as will
be discussed in the last section of the paper.
\end{enumerate}

\paragraph{The Main Results of the Paper.} In this paper we
show that the geodesic spray of the right invariant $H^1$ metric on
compact manifolds with boundary is smooth and has a unique flow which
corresponds to the solution of the averaged Euler equations
(\ref{mme.equations}), the correspondence given by right translation.
This has consequences that are similar to those for the usual Euler
equations, namely, local existence and uniqueness, smooth
dependence on initial condition and time, convergence of
Chorin-Marsden type product formulas, etc.
\medskip

Using this set up, we establish one of the main corollaries of
our results; namely, for the viscous analog of the
averaged Euler equations (\ref{mme.equations}),
\begin{equation}
\label{nsea}
\frac{\partial v^i}{\partial t} - \nu \triangle v^i
+ u^j \frac{\partial v^i}{\partial x^j}
-\alpha^2 \left[
\frac{\partial u^j}{\partial x^i} \right]  \triangle u_j
= - \frac{\partial p}{\partial x ^i}\,,
\end{equation}
which we call the averaged Navier-Stokes equations, or
the Navier-Stokes-$\alpha$ equations, we prove that
\begin{quotation}
{\it The solutions for the
corresponding viscous problem converge to those for the ideal
problem, as the viscosity goes to zero, even in the presence of
boundaries on uniform time intervals $[0,T]$, for $T > 0$,
independent of the viscosity. The size of the interval $[0,T]$ is governed
by the time of existence for the averaged Euler equations with the initial
data fixed.}
\end{quotation}
The inclusion of boundaries is a major difference from the situation with
the usual Navier-Stokes equations and the Euler equations, for which
such convergence is believed to not hold near boundaries because of the
generation of vorticity at the boundary. (See, for example, Marsden, Ebin
and Fischer [1972] and Chorin, Hughes, Marsden, and McCracken [1978]
for discussions).

\paragraph{Product Formulas.} Once one has the above geometric
setting and the smoothness of the spray is established, one can
``read off'' a number of interesting consequences. One of these is
a product formula stated in the following equation. This is one of
several possible product formulas, some of which are useful in
computational settings. The product formula we have in mind is the
mean motion version of a formula for the Euler equations having its
origins in Chorin [1969] and stated, along with a very simple proof
(based on the smoothness of the Lagrangian representation of the
equations) in Marsden, Ebin and Fischer [1972]. This formula
 is
\[
E _t = \lim_{n \rightarrow \infty} \left(P^\alpha\circ G _{t/n}
\right) ^n,
\]
where $E_t$ is the flow on the space of divergence free velocity
fields $u$ of the averaged Euler equations (with, say, zero
boundary conditions),  $P^\alpha$ is the $H ^2$-orthogonal
projection onto the divergence free vector fields zero on the
boundary (defined below), and where
$G_t$ is the unconstrained $H^1$ spray---that is,
the problem with the incompressibility condition dropped.

If there is no boundary, then $P^\alpha$ is the same as
$P$, the usual Hodge $L^2$ orthogonal projection. For the case of
zero slip boundary conditions, we have
\[
P^\alpha
= ( 1 - \alpha^2 \triangle) ^{-1} P  ( 1 - \alpha^2 \triangle)
\]
where the domain of $( 1 - \alpha^2 \triangle)$ is $H^2(TM) \cap
H^1_0(TM)$ (the Sobolev space of $H ^2$ vector fields vanishing on the
boundary). Notice that $P^\alpha$ is idempotent, so that it is a projection
in $H^2$ since
\[
\langle (1- \alpha^2\triangle) P^\alpha u, (1-
\alpha^2\triangle) u
\rangle_0
\ge 0\, ,
\]
where $\langle \cdot, \cdot
\rangle_0 $ is the $L^2$ inner product. Note, however, that the $H^1$ inner
product of $(1-\alpha^2\triangle)^{-1}\text{grad}\ p$ with divergence free
vectors $U$ in $H ^s \cap H _0^1$ vanishes in the sense that
\[
\langle (1-\alpha^2\triangle)^{-1} \text{grad}\ p, U \rangle_1
:=
\langle (1-\alpha^2 \triangle) (1- \alpha^2\triangle)^{-1}
\text{grad}\ p , U
\rangle_0 = \langle \text{grad}\ p , U \rangle_0 = 0.
\]

In the case of the viscous version of the equations, it is very
interesting that now {\it one does not require the vorticity creation
operator to correct the boundary term} (see also Marsden [1973, 1974],
Chorin, Hughes, Marsden and McCracken [1978] and Benfatto and Pulvirenti
[1986]). The form of this product formula is
\begin{align*}
F _t & = \lim_{n \rightarrow \infty} \left(S_{t/n} \circ E_{t/n}
\right) ^n\\
& = \lim_{n \rightarrow \infty} \left(S_{t/n} \circ P^\alpha \circ
G_{t/n}
\right) ^n
\end{align*}
where $S _{t} $ is the Stokes-$\alpha$ flow.

\paragraph{Smoothness of the Spray.} A crucial central result for
this paper is the smoothness of the spray of the averaged Euler
equations. Concretely, when we refer to the spray, we mean the
infinite dimensional vector field governing the dynamics written in
Lagrangian representation.  It is this vector field, which is shown
to be smooth in the standard sense of smooth vector fields on
infinite dimensional manifolds.

Smoothness of the spray has the direct consequence that the
averaged Euler equations are well-posed using the dynamical variable
$u$, for short time in the $H ^s$ topology for $s > n/2 + 1$. Note
that the vector field $v = (1 - \alpha ^2\triangle) u$ is
only in class $H ^{s - 2}$ and need not be $C ^0$. Smoothness also
plays a fundamental role for the derivation of the consequences
already mentioned, such as the limit of zero viscosity and the
product formulas.

Some additional results which are a consequence of the
smoothness of the spray are also worth stating. For example, if
$\eta$ is a volume preserving diffeomorphism close to the
identity, then there is a  unique initial velocity field whose
averaged Euler flow reaches $\eta$ in time one (and stays in a
neighborhood of the identity diffeomorphism). In addition,
even if the initial condition is only $H^s$, the flow is
infinitely differentiable in time. This latter fact is
surprising even for the usual Euler equations.\footnote{As was pointed
out by T. Kato (in an unpublished manuscript), this fact for the Euler
equations was rediscovered by several authors who were evidently unaware
that it is an immediate consequence of Ebin and Marsden
[1970].}

We conclude by outlining the general strategy for the proof of
smoothness of this spray as well as stating a number of related
results that follow from our method of proof.

\begin{enumerate}
\item The spray of the Camassa-Holm equation on the circle
is smooth on the tangent bundle of the $H ^s$ diffeomorphism
group of the circle, for $s > 3/2$. This follows from the fact
that the spray has no loss of derivatives (see Remark 3.5 of
Shkoller [1998] and Kouranbaeva [1999]) and the method of proof
in the present paper. We also note that the noncompact case
(replace the circle with the real line) should follow using
weighted Sobolev spaces, as in Cantor [1975].

\item More generally, the right invariant $H^1$ metric on the
full $H^s$ diffeomorphism group for $s > n/2 + 2$ of a compact
Riemannian $n$-manifold possibly with boundary and with
pointwise fixed boundary conditions, has a smooth spray. Using
the Moser [1966] estimates on products, we conjecture that the condition on
$s$ can be weakened to $s > n/2 + 1$.

\item The right translated projection $P^\alpha$ as a map from
the tangent bundle of the full diffeomorphism group (restricted to the
volume preserving ones) to the tangent bundle of the volume
preserving diffeomorphism group is smooth.

\item The sprays on the group of volume preserving
diffeomorphisms and that on the full group of diffeomorphism
are related by the derivative of the right translated
projection $P ^\alpha$.

\item Combining the last two facts, one arrives at smoothness of
the spray of the $H ^1 $ metric on the volume preserving
diffeomorphism group in the $H^s$ topology, and a careful
examination of the proof shows that $s > n/2 + 1$ is sufficient.

\end{enumerate}

\section{Diffeomorphism Groups}
In this section we set up the relevant groups of
diffeomorphisms that we shall need to study the averaged Euler
equations in Lagrangian representation.

\paragraph{Sobolev Spaces of Mappings.} Let $({M}, \langle \cdot ,
\cdot \rangle )$  be a compact oriented $C^\infty$ $n$-dimensional
Riemannian manifold possibly with boundary,
and let $({Q}, \langle \cdot, \cdot \rangle^Q)$ be a
$p$-dimensional compact Riemannian manifold without boundary.
By Sobolev's embedding theorem,
when $s > n/2 + k$, the set of Sobolev mappings $H^s(M,Q)$
 is a subset of $C^k(M,Q)$ with continuous inclusion, and so
for $s>n/2$, an $H^s$-map of $M$ into $Q$ is pointwise
well-defined.  Mappings in the space $H^s(M,Q)$ are those whose
first $s$ distributional derivatives are square integrable in {\it
any} system of charts covering the two manifolds.

For $s>n/2$, it is known that the space $H^s(M,Q)$ is a $C^\infty$
differentiable Hilbert manifold (see Palais [1968], Ebin and
Marsden [1970] and references therein). Let exp$:TQ
\rightarrow Q$ be the exponential mapping associated with $\langle
\cdot ,\cdot \rangle^Q$.  Then for each $\phi \in H^s(M,Q)$, the
map $\omega_{\operatorname{exp}}:T_\phi H^s(M,Q) \rightarrow
H^s(M, Q)$ is used to provide a differentiable structure which
is independent of the chosen metric, where
$\omega_{\operatorname{exp}}(v) = \operatorname{exp} \circ v$.

\paragraph{Diffeomorphism Groups.} The set of $H^s$ mappings from
$M$ to itself is not a smooth manifold; however, if we embed $M$
in its double $\tilde{M}$, then the set $H^s(M,\tilde{M})$ is a
$C^\infty$ Hilbert manifold, and  for $s> n/2+1$, we may form the
set $\mathcal{D}^s(M)$ consisting of $H^s$ maps $\eta$ mapping $M$
to $M$ with $H^s$ inverses. This space {\it is} a smooth manifold.
It is a well-known fact that the diffeomorphism group
$\mathcal{D}^s(M)$ is a
$C^\infty$ topological group for which the left translation
operators are continuous and the right translation operators are
smooth (Ebin and Marsden [1970] and references therein). One also
knows that $\eta: M
\rightarrow M$ has an extension to an element of (the connected
component of the identity of) $\mathcal{D}^s(\tilde{M})$ if and
only if $\eta$ lies in (the connected component of the identity
of) $\mathcal{D}^s(M)$.

Let $\mu$ be the Riemannian volume form on $M$, and denote by
\[
\mathcal{D}_\mu^s(M) := \{\eta \in \mathcal{D}^s(M) \mid
\eta^*(\mu)=\mu \}
\]
 the subgroup of $\mathcal{D}^s(M)$ consisting
of all volume preserving diffeomorphisms of class $H^s$.  For each
$\eta\in \mathcal{D}_\mu^s(M)$, we may use the $L^2$ Hodge
decomposition to define the projection
$P_\eta:T_\eta\mathcal{D}^s(M) \rightarrow T_\eta
\mathcal{D}_\mu^s(M)$ given by
$$ P_\eta (X) = (P_e(X \circ \eta^{-1}))\circ \eta, $$
where $X \in
T_\eta\mathcal{D}_\mu^s(M)$, and  $P_e$ is the $L^2$ orthogonal
projection onto the divergence-free vector fields on $M$. Recall
that this projection is given by
$$
P_e(v) = v - \operatorname{grad}p(v)- \operatorname{grad} b(v),
$$
where $p$ is the solution of the boundary value problem
\begin{align*}
\triangle p(v) &= \text{div}\  v  \quad \text{ in } M \\
p(v) & =0  \; \; \, \qquad \text{ on } \partial M,
\end{align*}
and $b$ solves
\begin{align*}
\triangle b(v) &= 0 \qquad \qquad\qquad \; \; \, \text{ in } M \\
\langle \operatorname{grad}  b(v), n \rangle   & = \langle v -
\operatorname{grad} p, n \rangle \quad \text{ on } \partial M,
\end{align*}
where $n$ is the orientation preserving normal vector field on $\partial M$.
The function $p$ is the pressure associated with $v$, while the function $b$
is a smooth extension
of the normal component of $v$ along $\partial M$ to the interior of $M$.
Subtraction of grad $b(v)$ is necessary as volume preserving
diffeomorphisms of
a manifold with boundary leave the boundary invariant.

\paragraph{Diffeomorphisms Leaving the Boundary
Pointwise Fixed.} We define the set
\[
\mathcal{D}_{\mu,{\rm fix}}^s(M)
= \{ \eta \in \mathcal{D}_\mu^s(M) \mid
\eta(x) =x \text{ for all } x \in \partial M\},
\]
that is, the volume preserving diffeomorphisms that leave the
boundary pointwise fixed.

\begin{thm}[Ebin and Marsden (1970)]\label{thm_em}
$\mathcal{D}_{\mu,{\rm fix}}^s(M)$ is a smooth subgroup of
$\mathcal{D}^s(M)$ with  Lie algebra $T_e\mathcal{D}_{\mu, {\rm
fix}}^s(M)$ consisting of the space of divergence free $H^s$
vector fields that vanish on $\partial M$.
\end{thm}
In other words, this group corresponds to the choice of the no
slip boundary conditions $u = 0$ for the averaged Euler
equations.

We can now define $P^\alpha_\eta : T_\eta \mathcal{D}^s(M) \rightarrow
T_\eta \mathcal{D}_{\mu,{\rm fix}}^s(M)$ by
\begin{equation}\label{pa}
P^\alpha_\eta(X) = \left[ (1-\alpha^2\triangle)^{-1} P_e (1 -
\alpha^2\triangle)
(X \circ \eta^{-1})
\right] \circ \eta.
\end{equation}

\paragraph{Alternative Boundary Conditions.} As is described
in Holm, Marsden and Ratiu [1998a], there is an alternative
choice of boundary conditions for the averaged Euler equations
that are more like normal boundary conditions than no slip
conditions. We will describe these conditions briefly and the
geometry and analysis of this situation is explored in
detail in Marsden, Ratiu and Shkoller [1999a].

Let $n $ denote the outward unit normal vector field along the
boundary of $M$ and let $N$ denote the corresponding normal bundle. We set
\[
\mathcal{D}_{\mu,{\rm normal}}^s(M)
= \{ \eta \in \mathcal{D}_\mu^s(M)
\mid T\eta |_{\partial M}: N \rightarrow N\}.
\]

This group $\mathcal{D}_{\mu,{\rm normal}}^s(M)$  corresponds
to an alternative choice of boundary conditions, namely
those divergence free vector fields $u$ that are tangent to
$\partial M$  and satisfy the boundary condition
\begin{equation}\label{sff}
\langle \nabla_n u, v \rangle = H _n(u, v) = \left\langle S _n (u), v
\right\rangle,
\end{equation}
at points of $\partial M$, for all divergence free vector
fields $v$ that are tangent to the boundary, where $H _n$ (or equivalently
$S _n$) denotes the second fundamental form of the boundary.

\section{Mean Hydrodynamics on the Subgroup
$\mathcal{D}_{\mu, {\rm fix}}^s(M)$}
\label{sec_spray}

\paragraph{$H^1$ Metric on $\mathcal{D}_{\mu, {\rm fix}}^s(M)$.}
In this section, we shall consider geodesic motion of the weak $H^1$ right
invariant metric on  the group $\mathcal{D}_{\mu,{\rm fix}}^s(M)$ which is
defined as follows. For $X,Y \in T_e \mathcal{D}_{\mu,{\rm fix}}^s(M)$, we
set
\begin{equation}
\label{s1}
\langle X, Y \rangle_{1}
= \int_M \left( \langle X(x),Y(x) \rangle +
\alpha^2\langle \nabla X(x), \nabla Y(x) \rangle \right) \mu(x),
\end{equation}
and extend $\langle \cdot, \cdot \rangle_{1}$ to $\mathcal{D}_{\mu, {\rm
fix}}^s(M)$ by right invariance.

\paragraph{Submanifold Geometry.}
Theorem \ref{thm_em} ensures that $\mathcal{D}_{\mu,{\rm fix}}^s(M)$  is
submanifold of $\mathcal{D}^s(M)$ in the strong $H^s$ topology. With the
induced metric (\ref{s1}), the submanifold $\mathcal{D}_{\mu,{\rm
fix}}^s(M)$ also inherits weak $H^1$  structure.  Our definition of the
projection $P^\alpha$ in (\ref{pa}) from tangent spaces of
$\mathcal{D}^s(M)$ to tangent spaces of $\mathcal{D}_{\mu,{\rm fix}}^s(M)$
is orthogonal with respect to this weak $H^2$ structure,  but $P^\alpha$
is not a Hodge projection.  In order to obtain a Hodge projection which is
$H^1$ orthogonal, one redefines the weak $H^1$ metric in terms of the
de-Rham Laplacian which is related to the rough Laplacian
$\triangle = \operatorname{Tr}\nabla
\nabla$  by  an additional term involving the Ricci tensor on $M$ (see, for
example, Section 2.3 of Shkoller [1998]).

Using the de Rham Laplacian, Shkoller [1998] has obtained expressions for
the unique $H^1$ Riemannian connection on $\mathcal{D}_\mu^s(M)$ when
$M$ is a compact boundaryless manifold.  Using this formula, together with
submanifold geometry, he has proven that the weak $H^1$ curvature operator
on $\mathcal{D}_\mu^s(M)$ is a continuous trilinear map in the strong $H^s$
topology.  This immediately implies the existence of unique solutions to
the Jacobi equations along the geodesics of the $H^1$ Riemannian
connection, and hence establishes the foundations for a Lagrangian
stability analysis. All of this analysis extends trivially to manifolds
with boundary on the subgroup $\mathcal{D}_{\mu,\rm{fix}}^s(M)$ if one
replaces the $H^1$ Hodge projection with our projection $P^\alpha$.

\paragraph{Euler-Poincar\'{e} equations on $T_e\mathcal{D}_{\mu, {\rm
fix}}^s(M)$.}
The Euler-Poincar\'{e} theorem (for background, see, for example, Marsden
and Ratiu [1999] and Holm, Marsden and Ratiu [1998a]) uses an additional
fact coming from the group structure of the problem, namely that the right
translation maps are smooth on
$\mathcal{D}_{\mu, {\rm fix}}^s(M)$ which can be used to translate geodesic
motion over the entire topological group into motion in the ``Lie algebra''
$T_e\mathcal{D}_{\mu, {\rm fix}}^s(M)$. We shall state this theorem in this
context.

\begin{thm}[Euler-Poincar\'{e}]\label{thm_ep}
Equip $\mathcal{D}_{\mu, {\rm fix}}^s(M)$ with the right invariant metric
$\langle \cdot, \cdot\rangle_{1}$.
Then, a curve $\eta(t)$ in
$\mathcal{D}_{\mu, {\rm fix}}^s(M)$ is a geodesic of this metric if and only if
$u(t) = T_{\eta(t)}R_{\eta(t)^{-1}} \dot \eta(t) = {\dot{\eta}}(t) \circ
\eta(t) ^{-1}$
satisfies
\begin{equation}\label{EP}
\frac{d}{dt} u(t) = - P^\alpha
\circ \operatorname{ad}^*_{u(t)}u(t)
\end{equation}
where $\operatorname{ad}^*_u$ is the formal adjoint of
$\operatorname{ad}_u$ with respect to the metric $\langle \cdot,
\cdot
\rangle_{1}$ at the identity, i.e.,
\[
 \langle \operatorname{ad}^*_uv,w \rangle_{1} = \langle v, [u,w]
\rangle_{1}
\]
for all $ u,v,w \in T_e \mathcal{D}_{\mu, {\rm fix}}^s(M)$.
\end{thm}

Next, we shall prove that a unique continuously differentiable geodesic
spray exists for the right invariant  $H^1$ metric obtained by right
translating $\langle \cdot, \cdot \rangle_{1}$ over the entire group, but
we note that even if a given metric does not have a $C^1$ spray from $H^s$
into $H^s$, but there is still an existence theorem for geodesics, then the
theorem is still true, when appropriately interpreted.

We note here that use of the right invariant $H^1$ metric (as opposed to
the {\it usual}, or naive, $H^1$ metric) is essential. For example, denote
by ${\mathcal M}^1$ (${\mathcal M}^0$) the manifold of $H^s$ maps from
${\mathbb T}^n$ into ${\mathbb T}^n$ with the weak $H^1$ ($L^2$) topology
induced by the usual $H^1$ ($L^2$) metric.  On the orthogonal complement of
Ker$(1-\triangle)$, ${\mathcal M}^1$ is totally geodesic in ${\mathcal
M}^0$, and so the geodesic equations for both metrics are the same.

Now, in the context of the Euler-Poincar\'{e} theorem we have just stated,
a straightforward computation of the geodesic spray of the right invariant
metric restricted to right invariant vector fields on $\mathcal{D}_{\mu,
{\rm fix}}^s(M)$ shows that the spray coincides with the equations
expressed in terms of the coadjoint action (of the group
$\mathcal{D}_{\mu,{\rm fix}}^s(M)$ on the dual of its Lie algebra).
Equivalently, there are two natural connections given by the Riemannian
structure and right translation, and the Euler-Poincar\'{e} equations can
be obtained from subtraction of the latter from the former.
 For the general Euler-Poincar\'{e} theorem for arbitrary Lagrangians, see
Theorem 5.1 and the Appendix of Bloch, Krishnaprasad, Marsden, and Ratiu
[1996] and Holm, Marsden and Ratiu [1998a] for the semidirect product
theory.

The form (\ref{EP}) is obtained by expressing the variations $\delta u(t)$
of curves $u(t) \in T_e\mathcal{D}_{\mu, {\rm fix}}^s(M)$  in terms of the
variations $\delta \eta(t)$ of curves $\eta(t)$ in
$\mathcal{D}_{\mu,{\rm fix}}^s(M)$. In particular let $\eta: U
\subset {\mathbb R}^2 \rightarrow \mathcal{D}_{\mu,{\rm fix}}^s(M)$ be a
smooth map  and let
\[
u (t,s) = T_{\eta(t,s)}R_{\eta(t,s)^{-1}}
\left[ \frac{\partial \eta(t,s) }{\partial t } \right]
\]
 and
\[
\xi (t,s) = T_{\eta(t,s)}R_{\eta(t,s)^{-1}}
\left[ \frac{\partial \eta(t,s) }{\partial s} \right].
\]
Then
$(\partial u/ \partial s)
- (\partial \xi / \partial t) = [u,\xi ]$ and we
obtain
$$ \delta u(t) = \dot{\xi} - \operatorname{ad}_u\xi.$$

In the case of the right invariant $H^1$ metric on
$\mathcal{D}_{\mu, {\rm fix}}^s(M)$, a straightforward computation (as in
Holm, Marsden and Ratiu [1998a]) shows that
\begin{equation}\nonumber
\operatorname{ad}^*_uu = (1- \alpha^2\triangle)^{-1}\left[
\nabla_{u(t)} v(t) - \alpha^2[\nabla u(t)]^t \cdot \triangle u(t)\right],
\end{equation}
where
$$ v = (1- \alpha^2 \triangle)u.$$
Thus, the Euler--Poincar\'e equation $\dot u = - P^\alpha
\operatorname{ad}^*_uu$ takes the form
\begin{equation}\label{ea}
\dot{v}(t) + \nabla_{u(t)} v(t)
- \alpha^2[\nabla u(t)]^t \cdot \triangle u(t)=
- \operatorname{grad} p(t)
\end{equation}
together with the divergence constraint $\operatorname{div}
u=0$, the no slip boundary conditions $u = 0$ on the boundary
$\partial M $, and the initial conditions
$u(0)=u_0.$
We call this equation the {\bfi averaged Euler equations\/} or
the {\bfi Euler-$\alpha$ equations\/}.

In components, and with the standard summation conventions and choice
of Levi-Civita connection and rough Laplacian, these equations read as
follows
\[
\frac{\partial v ^i}{\partial t }
+ \frac{\partial v ^i}{\partial x ^j} u ^j + \Gamma ^i_{kj}v ^k u ^j
- \alpha ^2
\left[ g ^{il}
\left(\frac{\partial u ^j}{\partial x ^l} + \Gamma ^j_{kl} u ^k\right)
(\triangle u)_j
\right] = - g ^{ij}\frac{\partial p }{\partial x ^j},
\]
where
\[
(\triangle u)_j =
\left(\delta^l_i \frac{\partial}{\partial x^j} + \Gamma ^l_{ij} \right)
\left(\frac{\partial u ^i}{\partial x^l} + \Gamma ^i_{ml} u ^m\right).
\]
Being Euler--Poincar\'e equations, of course these equations share
all the properties given by the general theory, such as a
Kelvin-Noether theorem, a Lie--Poisson Hamiltonian structure and
so on.

\paragraph{The geodesic spray on
$\mathcal{D}_{\mu,{\rm fix}}^s(M)$.} Now we are ready to state and
prove the first main result of the paper.
\begin{thm}
\label{thm_spray1}
For $s > n/2 + 1$, there exists a unique continuously differentiable
geodesic spray of the metric $\langle \cdot ,
\cdot \rangle_{1}$  on the group $\mathcal{D}_{\mu,{\rm fix}}^s(M)$.
\end{thm}
{\bf Proof.}  We compute the first variation of the
action function
\[
{\mathcal E}(\eta) = {\frac{1}{2}}\int_{\mathbb R} \langle
{\dot{\eta}}(t),  {\dot{\eta}}(t) \rangle_{1} dt,
\]
 which we
decompose as
\[
{\mathcal E}^0(\eta) = \frac{1}{2} \int_{\mathbb R} \langle
{\dot{\eta}}, {\dot{\eta}}\rangle_0 dt
\]
and
\[
{\mathcal E}^1(\eta)= \frac{\alpha^2}{2 }\int_{\mathbb R}
\langle
\nabla ({\dot{\eta}}
\circ \eta^{-1}), {\nabla (\dot{\eta}} \circ \eta^{-1}) \rangle_0
dt.
\]
We have
\begin{eqnarray*}
{\mathcal E}^1(\eta)
&=& {\frac{\alpha^2}{2}}
\int_{\mathbb R} \int_M \langle
\nabla({\dot{\eta}}\circ\eta^{-1})(y),
\nabla({\dot{\eta}}\circ\eta^{-1})(y)\rangle_y dy dt \\
&=& {\frac{\alpha^2}{2}} \int_{\mathbb R}
\int_M \langle \nabla {\dot{\eta}}(x)
\cdot [T\eta(x)]^{-1},\nabla {\dot{\eta}}(x)
\cdot [T\eta(x)]^{-1}\rangle_{\eta(x)} dxdt.
\end{eqnarray*}
Let $\epsilon \mapsto \eta^\epsilon$ be a smooth curve in
$\mathcal{D}_{\mu, {\rm fix}}^s(M)$ such that $\eta^0=\eta$ and
$(d /d\epsilon)|_{ \epsilon = 0} \eta^\epsilon = \delta
\eta$.  Then
\begin{eqnarray*}
\mathbf{d} {\mathcal E}^1(\eta)\cdot \delta \eta
&=& \alpha^2\int_{\mathbb R}\int_M
\left\langle
\left.\frac{D}{d\epsilon}\right|_0
\{\nabla {\dot{\eta}}^\epsilon \cdot
[T\eta^\epsilon]^{-1}\},
\nabla {\dot{\eta}}
\cdot [T\eta]^{-1}\right\rangle_{\eta(x)}dx dt\\
&=& \alpha^2\int_{\mathbb R}\int_M \left[ \left\langle
\left.\frac{D}{d\epsilon}\right|_0
\{\nabla_{[T\eta]^{-1} \langle \cdot \rangle}
{\dot{\eta}}^\epsilon ,\nabla
{\dot{\eta}}\cdot [T\eta]^{-1}\right\rangle_{\eta(x)}\right.\\
&&\quad \left. {\phantom{\frac{D^2}{d\epsilon}}} - \left\langle
\nabla\delta\eta \cdot  [T\eta]^{-1},
( \nabla{\dot{\eta}} \cdot [T\eta]^{-1})^t
(\nabla{\dot{\eta}} \cdot [T\eta]^{-1})
\right\rangle  \right] dx dt.
\end{eqnarray*}

Now fix $x \in M$, and let $s\mapsto \sigma_s$ be a smooth curve
in $M$ such that $\sigma_0=x$, and $(d/ds)|_{ s = 0} \sigma_s
= [T\eta(x)]^{-1} \cdot a$ for some $a \in
T_{\eta(x)}M$.  Define the parameterized surface
$(s, \epsilon) \mapsto f(s,\epsilon) :=
\eta^\epsilon \circ \sigma_s$.
Denote $(d/d \epsilon)(\cdot)$ by $\dot{({\cdot})}$ and
$(d/ds)(\cdot)$ by $(\cdot)'$.
Then,
\[
f(0,0)=\eta(x), \quad f(0,\epsilon)=\eta^\epsilon(x), \quad
f(s,0)=\eta(\sigma_s),
\]
and
\[
f'(s,\epsilon)= T\eta^\epsilon \cdot
\sigma_s', \quad \dot f(s,\epsilon) = (d/d \epsilon),
\eta^\epsilon(\sigma_s),
\]
so that
\[
 f'(0,0)=a, \text{ and }\dot f(0,0)=
\delta \eta(x).
\]
Hence, for any $a \in T_{\eta(x)}M$, and using $L\langle a \rangle$
to denote linear operation on $a$ by the linear operator $L$,
\[
\left. \frac{D}{d\epsilon} \right|_0
\nabla_{ [T\eta(x)]^{-1}\langle a \rangle }
= \left. \frac{D}{d\epsilon} \right|_0
\nabla_{\sigma'(0)} =\left. \frac{D}{d\epsilon} \right|_0
\left. \frac{D}{ds} \right|_0,
\]
and since
$$
\left. \frac{D}{d\epsilon} \right|_0
\left. \frac{D}{ds} \right|_0 {\dot{\eta}}^\epsilon =
\left. \frac{D}{ds} \right|_0
\left. \frac{D}{d\epsilon} \right|_0 {\dot{\eta}}^\epsilon
+R(\dot f, f')|_{0,0} {\dot{\eta}}^\epsilon,
$$
where $R$ is the curvature tensor on $M$, we get
\begin{eqnarray*}
\left. \frac{D}{d\epsilon} \right|_0
\nabla_{[T\eta(x)]^{-1} \langle a
\rangle } {\dot{\eta}}^\epsilon
&=& \nabla_{[T\eta(x)]^{-1} \langle a \rangle  }
\left. \frac{D}{d\epsilon} \right|_0 {\dot{\eta}}^\epsilon
+R(\dot f , f')|_{0,0} {\dot{\eta}} \\
&=& \nabla_{[T\eta]^{-1} \langle a  \rangle }
\frac{D}{dt} \delta \eta
+R(\delta \eta, a ) {\dot{\eta}}.
\end{eqnarray*}
Thus,
\begin{eqnarray*}
\mathbf{d} {\mathcal E}^1(\eta)
\cdot \delta \eta &=& \alpha^2\int_{\mathbb R}\int_M \left[
\langle \nabla [ (\nabla/dt) \delta \eta] \cdot [T\eta]^{-1} +
R(\delta \eta, \cdot)
{\dot{\eta}}, \nabla\cdot [T\eta]^{-1} \rangle \right.\\
&& \qquad \qquad
\left. \langle \nabla \delta \eta , (\nabla {\dot{\eta}} \cdot
[T\eta]^{-1})^t
\cdot (\nabla {\dot{\eta}}\cdot [T\eta]^{-1}) \cdot {[T\eta]^{-1}}^t \rangle
\right] dx dt.
\end{eqnarray*}
Now, using the same argument as above, we find that
\begin{eqnarray*}
&&\left\langle\nabla (\frac{D}{dt} \delta \eta)
\cdot [T\eta]^{-1}, \nabla
{\dot{\eta}} \cdot[T\eta]^{-1} \right\rangle \\
&&\qquad \qquad =
\left\langle \frac{D}{dt} \nabla \delta \eta
+ R(T\eta, {\dot{\eta}})\delta \eta,
\nabla{\dot{\eta}}\cdot [T\eta]^{-1}
\cdot {[T\eta]^{-1}}^t \right\rangle \\
&&\qquad \qquad =
-\left\langle \nabla \delta \eta,
\frac{D}{dt} \{ \nabla {\dot{\eta}} \cdot
[T\eta]^{-1}\cdot {[T\eta]^{-1}}^t\}\right\rangle +
\left\langle R({\dot{\eta}}, \cdot )
\nabla({\dot{\eta}} \circ \eta^{-1}), \delta\eta
\right\rangle.
\end{eqnarray*}
Finally,
\begin{eqnarray*}
\frac{D}{dt} \{ \nabla {\dot{\eta}}
\cdot [T\eta]^{-1} \cdot {[T\eta]^{-1}}^t\}
&=&
\nabla \frac{D}{dt} {\dot{\eta}}
\cdot [T\eta]^{-1} \cdot {[T\eta]^{-1}}^t
+R({\dot{\eta}}, T\eta){\dot{\eta}}
\cdot[T\eta]^{-1} \cdot {[T\eta]^{-1}}^t \\
&& - ( \nabla{\dot{\eta}}
\cdot [T\eta]^{-1}) \cdot (\nabla {\dot{\eta}} \cdot
[T\eta]^{-1}) \cdot {[T\eta]^{-1}}^t\\
&& -(\nabla{\dot{\eta}}\cdot [T\eta]^{-1})
\cdot (\nabla {\dot{\eta}} \cdot [T\eta]^{-1})^t
\cdot {[T\eta]^{-1}}^t.
\end{eqnarray*}
Integrating by parts, noting that the boundary terms vanish by
virtue of the subgroup
$\mathcal{D}_{\mu, {\rm fix}}^s(M)$, and denoting by $\nabla^*$ the $L^2$
formal adjoint of
$\nabla$, we have that
\begin{eqnarray*}
\mathbf{d} {\mathcal E}^1(\eta) \cdot \delta \eta
& = & \alpha^2\int_{\mathbb R} \int_M
-\left\langle \delta\eta, \nabla^*
\{\nabla (\nabla/dt) {\dot{\eta}} \cdot
[T\eta]^{-1} \cdot {[T\eta]^{-1}}^t\}\right.\\
&& \qquad  + \nabla^*\left[ (\nabla{\dot{\eta}} \cdot
[T\eta]^{-1})^t \cdot (\nabla{\dot{\eta}}
\cdot [T\eta]^{-1}) \cdot {[T\eta]^{-1}}^t
\right. \\
&& \qquad \left.-(\nabla{\dot{\eta}} [T\eta]^{-1}) \cdot
(\nabla {\dot{\eta}} [T\eta]^{-1})
\cdot {[T\eta]^{-1}}^t \right]\\
&& \qquad + \langle \operatorname{Tr}
R(\nabla {\dot{\eta}} \cdot
[T\eta]^{-1}, {\dot{\eta}})\cdot
+ \operatorname{Tr} R({\dot{\eta}}, \cdot
) \nabla {\dot{\eta}} \cdot [T\eta]^{-1}, \delta \eta
\rangle \\
&& \qquad - \langle \nabla^*
( R({\dot{\eta}}, \cdot ) {\dot{\eta}} \cdot {[T\eta]^{-1}}^t,
\delta \eta \rangle.
\end{eqnarray*}

Computing the first variation of ${\mathcal E}^0$, we obtain
\begin{eqnarray*}
\mathbf{d} {\mathcal E}^0(\eta) \cdot \delta \eta &=&
\int_{\mathbb R}\int_M
\left\langle \left.\frac{D}{d\epsilon}\right|_0 {\dot{\eta}}^\epsilon,
{\dot{\eta}}\right\rangle dx dt =
\int_{\mathbb R}\int_M
\left\langle \frac{D}{dt} \delta\eta,
{\dot{\eta}}\right\rangle dx dt \\
&=&
\int_{\mathbb R}\int_M
\left\langle -\frac{D}{dt} {\dot{\eta}},
 \delta \eta \right\rangle dx dt .
\end{eqnarray*}
Setting $\mathbf{d} {\mathcal E} \cdot \delta \eta = 0$, and
using the projector
$P^\alpha$ given by (\ref{pa}) gives
\begin{eqnarray}
&&P^\alpha_\eta \circ \nabla_{\dot{\eta}}{\dot{\eta}}=
P_\eta^\alpha \circ (1- \alpha^2\widehat{\triangle}_\eta)^{-1}
\left[
\nabla^*\left\{ -(\nabla{\dot{\eta}}
[T\eta]^{-1})^t (\nabla {\dot{\eta}}
[T\eta]^{-1}) \right.\right.\nonumber\\
&&\qquad \qquad \left. + \nabla{\dot{\eta}} [T\eta]^{-1}
\nabla{\dot{\eta}} [T\eta]^{-1}
 + (\nabla{\dot{\eta}}[T\eta]^{-1})
( \nabla{\dot{\eta}}[T\eta]^{-1})^t
\right\} {[T\eta]^{-1}}^t]\nonumber \\
&&\qquad \qquad -
 \{\operatorname{Tr}[ R(\nabla \dot{\eta}
T\eta^{-1} \langle \cdot \rangle,
\dot{\eta}) \cdot + R(\dot{\eta}, \cdot)\nabla
\dot{\eta} T\eta^{-1}
\langle \cdot \rangle] \nonumber\\
&&\qquad \qquad \left.+ \nabla^*\{ R(\dot{\eta},
{T\eta^{-1}}^t){\dot{\eta}} \} \right] ,
 \label{spray}
\end{eqnarray}
where
\begin{equation}
\label{hatlap.equation}
\widehat{\triangle}_\eta=-\nabla^*[\nabla(\cdot )
(T\eta)^{-1} {(T\eta)^{-1}}^t],
\end{equation}
and where again $T\eta^{-1}\langle \cdot \rangle$ defines
linear operation on a vector; explicitly, operating on a vector $V(x)$,
this is just shorthand notation for $[T\eta(x)]^{-1}\langle V(x)\rangle
= [T\eta(x)]^{-1} \cdot V(x)$.
Denote the quadratic form on the right-hand-side of
(\ref{spray}) by $P_\eta^\alpha \circ F^\alpha_\eta$.
Notice that $F_\eta^\alpha$ has no derivative loss and by standard
arguments is continuous from $H^s$ into $H^s$.  This is analogous
to the equation (3.7) of Theorem 3.3 in Shkoller [1998], so we
shall follow the proof of this theorem.  The following is a combination
of Lemmas 3.1 and 3.2 in Shkoller [1998], where the detailed proofs
can be found.

\begin{lemma} \label{l1}
Let
\[
{\widehat\triangle}_{(\cdot)} :  \cup_{\eta \in
\mathcal{D}^s_\mu(M)} H^s_{\eta, 0}(TM) | \mathcal{D}_{\mu,{\rm fix}}^s(M)
\longrightarrow \cup_{\eta \in
\mathcal{D}_{\mu,{\rm fix}}^s(M)} H^{s-2}_\eta(TM)
| \mathcal{D}_{\mu,{\rm fix}}^s(M)
\]
be defined by equation (\ref{hatlap.equation})
and be the identity on $\mathcal{D}_{\mu,{\rm fix}}^s(M)$.  Then
${\widehat\triangle}_{(\cdot)}$ is a $C^1$ bundle map.  Also, the
operator $(1- \widehat{\triangle}_{( \cdot )})^{-1}$ is a $C^1$
bundle map as well.
\end{lemma}

We also have the following lemma which is Lemma 4 of Appendix
A in Ebin and Marsden [1970].
\begin{lemma} \label{l2}
The $L^2$ Hodge projection $\eta \rightarrow P_\eta$ is smooth
as a function of $\eta$.
\end{lemma}

Choosing a local chart $({\mathcal U}, x^i)$ on $M$, we
may express the geodesic
spray ${\mathcal S}: T\mathcal{D}_{\mu,{\rm fix}}^s(M)
\rightarrow TT \mathcal{D}_{\mu,{\rm fix}}^s(M)$
in this chart by
\begin{eqnarray}
{\mathcal S}_\eta({\dot{\eta}}) = \frac{d}{dt}(\eta, {\dot{\eta}})
&=&
\left( \dot\eta, (\operatorname{Id}-P^\alpha_\eta)\circ
[ \nabla_{ \dot\eta \circ\eta^{-1} }
(\dot\eta\circ\eta^{-1})] \right. \nonumber\\
&&\left. - P^\alpha_\eta \left[ \Gamma_\eta({\dot{\eta}},{\dot{\eta}})
\right] + P^\alpha_\eta \circ F_\eta \right),
\label{local}
\end{eqnarray}
where
\[
\Gamma_\eta({\dot{\eta}}, {\dot{\eta}}) = \left\{
\Gamma^i_{jk} \left( {\dot{\eta}}^j \circ \eta^{-1} \right)
\left( {\dot{\eta}}^k
\circ \eta^{-1} \right)\frac{\partial}{\partial x^i}
\right\} \circ \eta.
\]

The fact that the projection $P^\alpha_\eta$ is $C^1$ in
$\eta$ follows from Lemma \ref{l1} and Lemma \ref{l2} since
we may write (\ref{pa}) as $P_\eta^\alpha(X)=
(1-\alpha^2\widehat{\triangle}_\eta)^{-1} P_\eta (1-\alpha^2
\widehat{\triangle}_\eta) X$.

Lemma \ref{l1} also shows that $F_\eta^\alpha$ is $C^1$ as well.
Hence, the term $P^\alpha_\eta \circ (F_\eta
-\Gamma_\eta({\dot{\eta}}, {\dot{\eta}}) )$ is $C^1$ in
$\eta$ from $H^s$ into $H^s$.

As for the remaining term,
\begin{align*}
(\text{Id} - P^\alpha_\eta) \circ
\nabla_{{\dot{\eta}}\circ\eta^{-1}}
({\dot{\eta}} \circ \eta^{-1}) & =
(1- \alpha^2\triangle)^{-1} \text{grad} \triangle^{-1}
(1- \alpha^2 \triangle)
\text{div}
\nabla_{{\dot{\eta}}\circ\eta^{-1}}
({\dot{\eta}} \circ \eta^{-1}) \\
& +
(1- \alpha^2 \triangle)^{-1} \text{grad} \triangle^{-1}
[\text{div}, (1- \alpha^2 \triangle)]
\nabla_{{\dot{\eta}}\circ\eta^{-1}}
({\dot{\eta}} \circ \eta^{-1}) .
\end{align*}
The fact that the first term on the right-hand-side is a $C^1$ map from
$H^s$ into $H^s$ follows identically the argument used by Ebin and Marsden
[1970] to prove smoothness of the spray of the right invariant $L^2$
metric on $\mathcal{D}_\mu^s(M)$.   The second term is $C^1$ by similar
reasoning as $[\text{div}, (1- \alpha^2\triangle)]$ is a
differential operator of order $2$.  This proves that
${\mathcal S}$ is
$C^1$.

The standard theorem for existence and uniqueness of ordinary
differential equations in a Banach space shows that unique solutions
locally exist and depend smoothly on initial conditions.
Q.E.D.
\bigskip

We remark that although the statement of Lemma \ref{l1} is
$C^1$, it should be clear that the $C^k$ result could be
obtained for any positive integer $k$ by (a messy) induction;
this type of reasoning led to the $C^\infty$ regularity of the
$L^2$ Hodge projection $P$ in Lemma \ref{l2}.

As we remarked earlier, the method of proof we have given for
the smoothness of the spray of the right invariant $H^1$ metric
on $\mathcal{D}_{\mu, {\text fix}}^s(M)$ also provides
well-posedness for the one dimensional Camassa-Holm equation on
${\mathbb S}^1$, since the spray in this case does not have a
loss of derivative (c.f. Remark 3.5 in Shkoller [1998]). The
situation of the present paper is, however, complicated in a
nontrival way by the presence of boundaries.  We state the
situation for the CH equation as the following theorem.
\begin{thm}\label{1dch.thm}
The Cauchy problem for the 1D CH equation (\ref{1dch.eqn}), given by
\begin{equation}
\ddot\eta = -\left[ (1-\partial_y^2)^{-1} \partial_y
\left((\dot\eta \circ \eta^{-1})^2 +
\frac{1}{2}(\dot\eta \circ \eta^{-1})^2_y \right) \right] \circ \eta
\end{equation}
with initial conditions
\[
\eta(0) =e, \quad  \dot\eta(0) = u_0,
\]
has a unique solution $(\eta, {\dot{\eta}})$ in
$\mathcal{D}^s({\mathbb S}^1)\times H^s({\mathbb S}^1)$ for $s>
{\frac{3}{2}}$ on a finite time interval where the solution has
$C^1$ dependence on time and smooth dependence on initial data.
\end{thm}

Theorem \ref{thm_spray1} implies many other results too, such
as  finite smoothness with respect to initial conditions in
Eulerian representation if one is willing to map between
different Sobolev spaces, smoothness with respect to the time
variable, etc.

\paragraph{Remark.} The subgroup $\mathcal{D}_{\mu, {\rm fix}}^s(M)$
ensures that the boundary terms vanish in the above variational
principle, as the variation $\delta \eta (x) = 0$ for all $x \in
\partial M$.  For variations which are not required to vanish on
the boundary, one obtains natural boundary conditions on
$\partial M$, which are discussed in Marsden, Ratiu and Shkoller [1999a].

\section{The limit of zero viscosity on $\mathcal{D}_{\mu, {\rm fix}}^s(M)$}

The {\bf Navier-Stokes $\alpha$-model} is obtained by adding viscous
diffusion to the  Euler-$\alpha$ model. The equations are given by
\begin{equation}\label{nse}
\partial_t u - \nu \triangle u
+ (1-\alpha^2\triangle)^{-1}\left[
\nabla_u(1-\alpha^2 \triangle)u
- \alpha^2 \nabla u^t \cdot\triangle u\right]=
 - (1-\alpha^2\triangle)^{-1} \operatorname{grad} \ p.
\end{equation}
Foias, Holm, and Titi [1999] have proven global well-posedness of
(\ref{nse}) and have obtained estimates on the dimension of the
global attractor.  Chen et al. [1999] have given some very
interesting results concerning the applicability of these equations to
high Reynolds number flows in channels, including relations to the law of
the wall.

Having proven the smoothness of the geodesic spray of the Euler-$\alpha$
equations, we follow Ebin and Marsden [1970]  and use a product formula
approach to prove the existence of the limit of zero viscosity of
(\ref{nse}).  In the case that
$\alpha=0$, this limiting procedure is believed to be valid only for
compact manifolds without boundary (e.g., for flows with periodic boundary
conditions), as the Navier-Stokes equations and the Euler equations do not
share the same boundary conditions on manifolds with boundary.  When,
$\alpha\neq 0$, however, as we shall discuss in the last section, a
certain type of elasticity is added into the Euler-$\alpha$ model, and the
mean motion of the fluid exhibits normal stress effects. Because of this,
we may prescribe zero velocity boundary conditions even in the inviscid
limit, and thus extend the limit of zero viscosity theorems for the
averaged Euler equations to manifolds with boundary.

The following is the Euler-$\alpha$ version of Theorem 13.1 of Ebin
and Marsden [1970].
\begin{thm}\label{product}
Let $S: T\mathcal{D}_{\mu, {\rm fix}}^s(M) \rightarrow TT \mathcal{D}_{\mu,
{\rm fix}}^s(M)$
be the
Euler-$\alpha$ vector field given by (\ref{local}). Let
$T:T_e\mathcal{D}_{\mu,{\rm fix}}^s(M)
\rightarrow T_e \dot \mathcal{D}_\mu^{s-\sigma}(M)$ be a given map, where
$\sigma \ge 2$.  Let $T$ be a bounded linear map which generates a
strongly-continuous
semi-group $F_t: T_e\mathcal{D}_{\mu,{\rm fix}}^s(M) \rightarrow T_e
\mathcal{D}_\mu^s(M)$,
$t\ge 0$, and satisfies $\|F_t\|_s \le e^{\beta t}$ for some $\beta > 0$
and some $s$.
Extend $F_t$ to $T\mathcal{D}_{\mu, {\rm fix}}^s(M)$ by
$$ \tilde{F}_t(X_\eta) = TR_\eta \cdot F_t \cdot TR_{\eta^{-1}} (X_\eta)$$
for $X_\eta \in T_\eta\mathcal{D}_{\mu, {\rm fix}}^s(M)$, and let
$\tilde{T}$ be the
vector
field $\tilde{T}:T\mathcal{D}_{\mu, {\rm fix}}^s(M) \rightarrow
TT\mathcal{D}_\mu^{s-\sigma}(M)$
associated to the flow $\tilde{F}_t$.

Then $S+\nu \tilde{T}$ generates a unique local uniformly Lipschitz flow on
$T\mathcal{D}_{\mu, {\rm fix}}^s(M)$ for $\nu \ge 0$, and the integral curves
$c^\nu(t)$ with
$c^\nu(0)=x$ extend for a fixed time $\tau >0$ independent of $\nu$ and are
unique.
Further,
$$ \lim_{\nu \rightarrow 0} c^\nu(t) = c^0(t)$$
for each $t$, $0\le t < \tau$, the limit being in the $H^s$ topology,
$s>(n/2)+1+2\sigma$.
\end{thm}

See Lemmas 2-5 in Appendix B of Ebin and Marsden [1970] for an explanation of
$2\sigma$ in the hypothesis of Theorem \ref{product}.
Since Theorem \ref{thm_spray1} proves that the geodesic spray is smooth on
$\mathcal{D}_{\mu, {\rm fix}}^s(M)$, we must verify the hypothesis of Theorem
\ref{product}
for $T= -\triangle$.

\begin{lemma}
The operator $\triangle$ generates a strongly continuous
contraction semi-group on the space $T_e \mathcal{D}_{\mu,{\rm
fix}}^s(M)$, for $s>(n/2)+1$.
\end{lemma}
{\bf Proof.}
Standard energy estimates give the result. See, for example,
Theorem 3.2 of Temam [1988], page 70, and Theorem 13.2 of Ebin
and Marsden [1970]. Q.E.D.

\paragraph{Remark.}
With initial data $u_0 \in T_e\dot
\mathcal{D}_{\mu, {\rm fix}}^\infty(M)$, the solution $u(t)$ is
also $C^\infty$ as consequence of the regularization of
parabolic flows.
\medskip

The way one proves the product formulas mentioned in the
introduction is exactly the same as is outlined in Ebin and
Marsden [1970] and Marsden, Ebin and Fischer [1972]; one uses
the general theorems on product formulas (formulas of
Lie-Trotter type) for smooth vector fields on infinite
dimensional manifolds, as discussed in, for example, Chorin,
Hughes, Marsden and McCracken [1978] and in Abraham, Marsden
and Ratiu [1988] using the smoothness of the spray and the
projection operator. For product formulas involving the Stokes
operator, one couples the smoothness of the spray with the
analytic property of the generated semigroup, as explained in
Ebin and Marsden [1970]. Of course this technique is closely
related to Theorem \ref{product} discussed above.

\section{Second-grade Fluids}
\label{SGF}
As noted in Chen et al. [1999], non-Newtonian fluids
of second grade satisfy the same type of equations as the mean
motion model equations for incompressible ideal fluids. In
fact, we shall show that in the incompressible case, the
equations are identical in the limit of zero viscosity. For simplicity we
shall make the presentation in Euclidean space. (See Marsden and Hughes
[1983] for how to put this discussion on manifolds).

The constitutive equation of fluids for a fluid of grade two is
given by (see, for example, Noll and Truesdell [1965])
  \begin{equation}
  \label{SGFLAW}
  \mathbf{T} = -\bar{p} \mathbf{1} + \nu \mathbf{A} _1 + \alpha_1
\mathbf{A} _2
   + \alpha_2 \mathbf{A} _1^2,
  \end{equation}
where $\mathbf{T}$ is the stress tensor, $\bar{p}$ is the
hydrostatic pressure, $\mathbf{1}$ is a unit tensor, $\nu$ is
viscosity, $\alpha_1$ and $\alpha_2$ are normal stress moduli.
Here, $\mathbf{A}_1$ and $\mathbf{A}_2$ are Rivlin-Ericksen tensors (Rivlin
and Ericksen [1955]) defined by
\begin{align*}
  \mathbf{A}_1 & =  \mathbf{\nabla} u + (\mathbf{\nabla} u)^t\, ,
\\
  \mathbf{A}_2 & = \frac{D\mathbf{A}_1}{dt}
+ (\mathbf{\nabla} u)^t
\mathbf{A}_1+\mathbf{A}_1\mathbf{\nabla} u\, .
\end{align*}
Here, the tensors are of type $(1, 1)$ and $D/dt = (\partial/
\partial t) + u \cdot \nabla$ is the material derivative.

To satisfy the Clausius-Duhem inequality, the first
material constant has to satisfy $\nu \geq 0$ and the constants
$\alpha _1 $ and $\alpha _2$ satisfy
$  \alpha_1+\alpha_2=0\, .
$
If in addition one asks that the specific Helmholtz free energy is
a minimum when the fluid is in equilibrium, then
  \begin{equation}\nonumber
  \alpha_1 \geq 0\, .
  \end{equation}
The results expressed by the last inequalities are derived in
Dunn and Fosdick [1974].

We now turn to the incompressible case,
i.e., $\operatorname{div} u = 0$. Under this assumption, we
make the following calculations on the stress tensor:
$ \operatorname{div} \mathbf{T}$ :
  \begin{eqnarray*}
  \operatorname{div}\mathbf{T} & = & -\operatorname{div}
\bar{p} \mathbf{1}+\nu\,\operatorname{div}\mathbf{A}_1
 + \alpha_1(\operatorname{div} \mathbf{A}_2
  -\operatorname{div}\mathbf{A}_1^2)\, , \\
  \operatorname{div}\mathbf{A}_1 & = &
\operatorname{div}(\mathbf{\nabla} u)^t = \triangle u\, ,
\\ \operatorname{div}\mathbf{A}_2
& = & \frac{\partial }{\partial
t}\operatorname{div}\mathbf{A}_1  +
          \operatorname{div} (( u\cdot\mathbf{\nabla}
)\mathbf{A}_1) \\
            & & +  \operatorname{div}
(\mathbf{A}_1\mathbf{\nabla} u ) +
\operatorname{div}
((\mathbf{\nabla} u)^t\,\mathbf{A}^1)\, ,\\
  \operatorname{div} (( u\cdot\mathbf{\nabla} )
    \mathbf{A}_1)
& = &
\mathbf{\nabla} u\cdot\mathbf{\nabla}\mathbf{A}_1 + u \cdot\mathbf{\nabla}
(\triangle u ) \, ,\\
  \operatorname{div}(\mathbf{A}_1\mathbf{\nabla} u ) & = &
\sum_k(\triangle u)^k\mathbf{\nabla} u^k +
\frac{1}{2}\mathbf{\nabla}
(\operatorname{Tr}(\mathbf{\nabla} u )^2)
   + \frac{1}{2}\mathbf{\nabla}
(\operatorname{Tr}(\mathbf{\nabla} u
(\mathbf{\nabla} u )^t))
\, ,\\
\operatorname{div}
((\mathbf{\nabla} u)^t\,
\mathbf{A}_1)&=&\triangle u \cdot\mathbf{A}_1
+
(\mathbf{\nabla} u)^t\cdot\mathbf{\nabla}\mathbf{A}_1\, , \\
  \operatorname{div}(\mathbf{A}_1^2
)&=&\triangle u\cdot\mathbf{A}_1 +
(\mathbf{\nabla} u)^t\cdot\mathbf{\nabla}\mathbf{A}_1 +
  \mathbf{\nabla} u\cdot\mathbf{\nabla}\mathbf{A}_1 \, ,\\
  \operatorname{div}\mathbf{A}_2
- \operatorname{div}(\mathbf{A}_1^2
)&=&\frac{\partial }{\partial
t}(\triangle u ) +
  u\cdot\mathbf{\nabla} (\triangle u ) \\ & & +
(\mathbf{\nabla} u)^t \cdot \triangle u +
\frac{1}{2}\mathbf{\nabla}
  (\operatorname{Tr}(\mathbf{\nabla} u )^2) +
\frac{1}{2}\mathbf{\nabla}
(\operatorname{Tr}(\mathbf{\nabla} u
(\mathbf{\nabla} u)^t)) \, .\\
  \end{eqnarray*}
Substituting  $\operatorname{div} \mathbf{T}$ into the balance
of momentum with no body force acting on the fluid,
  \begin{equation}\nonumber
  \rho \frac{D u}{dt}=\operatorname{div} \mathbf{T}\, ,
  \end{equation}
and making use of the calculations above, we obtain the following
equation for a second-grade fluid
\begin{equation}\label{second_grade.equation}
\frac{D}{dt}(1-\tilde{\alpha}_1\triangle ) u =\tilde{\nu
}\triangle u +
\tilde{\alpha}_1(\mathbf{\nabla} u)^t\cdot \triangle u +
\mathbf{\nabla} \left( \tilde{\alpha }_1\,
\mathbf{D}\colon\mathbf{D} - \frac{\bar{p} }{\rho } \right)\, ,
\end{equation}
  where
$\mathbf{D}\colon \!\mathbf{D} = \operatorname{Tr}(\mathbf{D}
\mathbf{D}^t)$ and
\[
  \tilde{\alpha}_1 = \frac{\alpha_1}{\rho}\, ,
\quad \tilde{\nu}=\frac{\nu}
  {\rho}\, ,
\]
  and the stretch tensor is given by
  \begin{eqnarray*}
  \mathbf{D} = \frac{1}{2}\mathbf{A}_1\, .
  \end{eqnarray*}

The theory of second-grade fluids has been developed over the
last 40 years by a number of people, such as Noll and
Truesdell  [1965], Coleman and Markovitz [1964], Markovitz and
Coleman [1964], and Coleman, Markovitz, and Noll [1966].
These authors have shown that second grade fluids can be
regarded (like the Navier-Stokes theory) as an approximation
to the general theory of simple fluids with fading memory,
valid in sufficiently slow flows.

\paragraph{Inviscid Second Grade Fluids.} Letting the viscosity $\nu =0$ in
(\ref{second_grade.equation}) and defining
$\alpha^2 =
  \tilde{\alpha }_1$ , $ v =
 u -\alpha^2\triangle u $ gives the following equation
\begin{equation}
  \label{SGFEQN}
  \frac{D v }{dt} - \alpha^2 (\mathbf{\nabla} u)^t
\cdot\triangle u
 = - \mathrm{grad}\ P\,
,\quad
 \operatorname{div} u  =\operatorname{div} v =
  0\, ,
\end{equation}
where the second-grade fluid pressure function is given by
\[
 P=\frac{\bar{p} }{\rho } -
  \frac{\alpha^2}{2}\operatorname{Tr}(\mathbf{\nabla} u
  \cdot (\mathbf{\nabla} u)^t) -
\frac{\alpha^2}{2}\operatorname{Tr}(\mathbf{\nabla}u)^2
\, ,
\]
and where  $P$ is the generalized pressure.

We observe that {\it the equations (\ref{SGFEQN}) are identical to the
averaged Euler equations (\ref{ea})}. In either case, one can
eliminate the pressure terms in the standard fashion using the
projection to the divergence free part or, if one prefers, one can view the
pressure as being determined implicitly from the incompressibility
constraint. While these equations are the same, the physical interpretation
and the basic mechanics are different.

\paragraph{Comments on the Limit of Zero Viscosity.}
The equations for a second grade fluid with viscosity
(\ref{second_grade.equation}) are {\it not} the same as the
Navier-Stokes$-\alpha$ equations (\ref{nsea}). The difference
is simply that in the case of second grade fluids, one uses
the Laplacian of the velocity vector $u$ while in the
Navier-Stokes$-\alpha$ equations one uses the Laplacian
of the momentum $v$.

The theory for second grade fluids is, in a sense, much easier than
that for the averaged Navier-Stokes equations. This is because when the
equations are written with the dynamical variable
$u$, there is no derivative loss in the dissipative term
and so the vector field in Lagrangian coordinates is already
smooth. {\it The result on the limit of zero
viscosity for second grade fluids corresponding to that
for the Navier-Stokes$-\alpha$ equations is elementary,
following simply from the smoothness of the flow of
a smooth vector field as a function of parameters.}

In the literature on second-grade fluids it is generally
assumed that the viscosity does not vanish. However following
the approximation procedure of Coleman and Markovitz [1964],
and Noll and Truesdell  [1965] one can show that the case of
zero viscosity does not contradict their theory.
Hence {\it for a simple fluid with fading memory, a second-grade fluid
with zero viscosity furnishes a complete second-order
correction, for normal stress effects, to the theory of perfect
(in the terminology of Noll and Truesdell) elastic fluids, in
the limit of slow motion.} Notice that $\alpha_1$  is a material constant
in second-grade fluid theory, whereas $\alpha$ is the fluctuation amplitude
scale parameter (i.e., a flow regime parameter) in the average Euler
equations.

Another aspect to be mentioned here is that the theory of
simple fluids obeys the basic mechanical principles: the
principle of material frame-indifference, which assert  that
the response of a material is the same for all observers, and
the principles of determinism and local action, which asserts
that the present stress at a particle is determined by the
history of an arbitrary small neighborhood of that particle.
The same principles are true of the averaged Euler equations
and the Navier-Stokes$-\alpha$ equations.

\paragraph{Existence and Uniqueness.} In the viscous case,
existence and uniqueness has already been shown by Cioranescu and
Gorailt [1997] and Galdi [1995], although as we have mentioned,
many results, such as short time existence and uniqueness and
the limit of zero viscosity results follow from our work too.

The question of the existence and dimension of attractors
for the Navier-Stokes$-\alpha$ equations is
addressed in Foias, Holm and Titi [1999]. We note that these
techniques do not allow one to take the limit of zero
viscosity, whereas ours do allow such a limit.

\paragraph{Specific Solutions.} Several formal solutions (ignoring boundary
conditions) have been given for planar homogeneous incompressible flows of
second grade fluids. The stream function formulation of the dynamics of such
flows in two dimensions, with stream function $\psi$ and
velocity
$u$ related by
\[
u = \left( \frac{\partial\psi}{\partial y}, -\frac{\partial\psi}{\partial
x}, 0\right),
\]
 is given by
\begin{equation}
\label{2ndgrd-dyn}
q_t + \nabla_uq = \nu \Delta \omega,
\end{equation}
where
\[
q=-(1-\alpha^2\Delta)\Delta\psi
\quad  \mbox{and} \quad \omega=-\Delta\psi .
\]
>For example, Ting [1963] provided a set of exact solutions for planar
startup flows. Further, Rajagopal and Gupta [1981]
showed that equation (\ref{2ndgrd-dyn}) has separable solutions that
may be interpreted as, e.g., a damped lattice of vortices,
two impinging rotational flows, and a viscoelastic analog to Kelvin's
``Cat's eyes'' vortices. These formal separable solutions impose the
condition
\begin{equation} \label{c-cond}
\Delta \psi = c\ \psi,
\end{equation}
where $c$ is a nonzero constant. Of course, this condition implies the
linear relation $q=-(1-\alpha^2c)\,c\,\psi$. Hence, the transport term
$\nabla_uq=J(q,\psi)$ vanishes in~(\ref{2ndgrd-dyn}) and the remaining
terms impose
\[
q_t= \frac{\nu{q}}{1-\alpha^2c} ,
\]
which implies exponential time
dependence for $q$ and, thus, for $\psi$. For example, the stream
function for the viscoelastic analog to Kelvin's ``Cat's eye'' vortices
found in Rajagopal and Gupta [1981] is given in separated form by
\begin{equation} \label{eye}
\psi = A\ \cosh (ax)\ \cos (by)\ \exp (\lambda t)\,.
\end{equation}
This stream function is a solution of~(\ref{2ndgrd-dyn}) under
condition~(\ref{c-cond}) for arbitrary real $A, a, b$, with $c=a^2-b^2$ and
$\lambda=\nu/(1-\alpha^2c)$. Relation ~(\ref{eye}) with $\lambda=0$ and
$A,a,b$  arbitrary is also a {\it stationary} solution of
equation~(\ref{2ndgrd-dyn}) in the limiting case that $\nu = 0$.

Two-dimensional incompressible Navier-Stokes-$\alpha$ dynamics may be
expressed in a form similar to equation~(\ref{2ndgrd-dyn}), namely,
\begin{equation}
\label{qdynamics}
q_t + \nabla_uq = \nu \Delta q,
\end{equation}
where
\[
q\equiv\hat{z}\cdot\hbox{curl}{v}=-(1-\alpha^2\Delta)\Delta\psi
\]
 is
potential vorticity. Thus, under condition~(\ref{c-cond}) (and still
ignoring boundary conditions), separable solutions for second-grade fluids
satisfying~(\ref{2ndgrd-dyn}) become separable solutions for the
two-dimensional Navier-Stokes-$\alpha$ model satisfying~(\ref{qdynamics})
under the change $\nu\to(1-\alpha^2c)\nu$. In the case $\nu=0$, formal
solutions satisfying the linear relation~(\ref{c-cond}) are common to
both theories. Another example of a steady flow that is common to both
theories in the case $\nu=0$ is the Gaussian jet, for which
$\psi=\exp(-y^2)$.

For the ideal limiting case in which $\nu=0$ it should be clear that
steady incompressible flows of {\it both} second grade fluids {\it and}
the two-dimensional Euler-$\alpha$ model occur whenever $q$ and
$\psi$ are functionally (not just linearly) related.  For example, for
periodic boundary conditions, critical points of the functional
\begin{eqnarray*}
H_{\Phi }=H+C_{\Phi }= \frac{1}{2}\int [q\psi+\Phi (q)]dxdy\ ,
\end{eqnarray*}
(the sum of the energy and the domain integral of an arbitrary function
$\Phi $ of the potential vorticity) are steady flows of both models.
This observation leads immediately to a unified viewpoint of nonlinear
stability criteria for steady periodic flows of either
model, following the energy-Casimir method, as in Holm, Marsden, Ratiu and
Weinstein [1985].

Exact solutions of the three-dimensional Navier-Stokes-$\alpha$ model may
also be sought by adapting the classical solutions for the two-dimensional
Navier-Stokes equations, e.g., for flows between two rotating disks,
Beltrami flows, etc., as reviewed, e.g., in Wang [1991].

\paragraph{Acknowledgments.}
We thank Peter Constantin, David Ebin, Ciprian Foias, Gerard Misiolek, and
Edriss Titi for valuable discussions and remarks. The research of JEM
was supported by the California Institute of Technology and National
Science Foundation Grant ATM-98-73133, and that of DDH, TSR, and SS  was
partially supported by DOE.

\subsection*{References}
\addcontentsline{toc}{section}{References}

\begin{description}

\item Abraham, R., J.E. Marsden, and T.S. Ratiu [1988]
{\it Manifolds, Tensor Analysis, and Applications.\/}  Second
Edition, Applied Mathematical Sciences {\bf 75},
Springer-Verlag.

\item Alber, M.S., R. Camassa, D.D. Holm and J.E. Marsden
[1994] The geometry of peaked solitons and billiard
solutions of a class of integrable pde's, {\it Lett. Math.
Phys.\/} {\bf 32}, 137--151.

\item Arnold, V.I. [1966]
Sur la g\'{e}om\'{e}trie differentielle
des groupes de Lie de dimenson
infinie et ses applications \`{a}
l'hydrodynamique des fluids parfaits.
{\it Ann. Inst. Fourier, Grenoble\/} {\bf 16}, 319--361.

\item Arnold, V.I. [1978]
{\it Mathematical Methods of Classical Mechanics.}
Graduate Texts in Math. {\bf 60}, Springer Verlag.
(Second Edition, 1989).

\item Arnold, V.I. and B. Khesin [1998]
{\it Topological Methods in Hydrodynamics.}
Appl. Math. Sciences {\bf 125}, Springer-Verlag.

\item Barenblatt, G.I. and A.J. Chorin [1998a]
New Perspectives in turbulence: scaling laws,
asymptotics, and intermittency, {\it SIAM Rev.},
{\bf 40}, 265--291.

\item Barenblatt, G.I. and A.J. Chorin [1998b] Scaling laws and
vanishing viscosity limits in turbulence theory. {\it Recent
advances in partial differential equations, Venice 1996,
Proc. Sympos. Appl. Math.}, {\bf }54, 1--25.

\item Benfatto, G. and M. Pulvirenti [1986] Convergence of the
Chorin-Marsden product formula in the half-plane {\it Comm. Math.
Phys.\/} {\bf 106}, 427--458.

\item Bloch, A., P.S. Krishnaprasad, J.E. Marsden, and T. Ratiu
[1996] The Euler Poincar\'{e} equations and double bracket
dissipation, {\it Comm. Math. Phys.} {\bf 175}, 1--42.

\item Camassa, R. and D.D. Holm [1993] An integrable
shallow water equation
with peaked solitons,  {\it Phys. Rev. Lett.\/}, {\bf 71},
1661-1664.

\item Camassa, R., D.D. Holm and J.M. Hyman [1994]
A new integrable shallow water equation.
{\it Adv. Appl. Mech.}, {\bf 31}, 1--33.

\item Cantor, M. [1975]
Perfect fluid flows over $R^n$ with asymptotic conditions.
{\it J. Func. Anal.\/} {\bf 18}, 73--84.

\item Chen, S.Y., C. Foias, D.D. Holm, E.J. Olson, E.S. Titi and S. Wynne
[1999] The Camassa-Holm equations as a closure model for turbulent channel
and pipe flow, {\it Phys. Rev. Lett.,} {\bf 81}, 5338--5341.

\item Chorin, A.J. [1969]  On the convergence of discrete
approximations to the Navier-Stokes equations. {\it Math. Comp.}
{\bf 23}, 341--353.

\item Chorin, A.J., T.J.R. Hughes, J.E. Marsden, and M. McCracken
[1978] Product Formulas and Numerical Algorithms,
{\it Comm. Pure  Appl. Math.\/} {\bf 31}, 205--256.

\item Cioranescu, D., and V. Girault, [1997]
Weak and classical solutions of a family of second grade fluids,
{\it Inter. J. Non-Linear Mech.} {\bf 32}, 317--335.

\item Coleman, B.D., and H. Markovitz, [1964] Normal stress
effects in second-order fluids, {\it J. Appl. Phys.} {\bf 35},
1--9.

\item Coleman, B.D., H. Markovitz and W. Noll, [1966]
{\it Viscometric flows of Non - Newtonian Fluids},
Springer-Verlag,  New York Inc.

\item Coleman, B.D., and W. Noll [1960] An approximation
theorem for functionals, with applications in continuum mechanics,
{\it Arch. Rat. Mech. Anal.} {\bf 6}, 355--370.

\item Dunn, J.E., and R.L. Fosdick, [1974] Thermodynamics,
stabilityand boundedness of fluids of complexity $2$ and fluids
of second grade, {\it Arch. Rat. Mech. Anal.} {\bf 56}, 191--252.

\item Ebin, D. and J.E. Marsden [1970] Groups of diffeomorphisms
and the motion of an incompressible fluid. {\it Ann. of Math.}
{\bf 92}, 102-163.

\item Foias, C., D.D. Holm and E.S. Titi [1999]
Global well-posedness for the three dimensional
viscous Camassa-Holm equation, {\it preprint}.

\item Galdi, G.P. [1995] {\it Mathematical theory of second-grade
fluids}, CISM Courses and Lectures, 344, Springer, Vienna.

\item Holm, D.D., J.E. Marsden and T.S. Ratiu [1998a]
The Euler--Poincar\'{e} equations and semidirect products
with applications to continuum theories,
{\it Adv. in Math.}, {\bf 137}, 1-81.

\item Holm, D.D., J.E. Marsden and T.S. Ratiu [1998b]
Euler-Poincar\'e models of ideal fluids with nonlinear dispersion,
{\it Phys. Rev. Lett.} {\bf 349}, 4173-4177.

\item Holm, D.D., J.E. Marsden and T.S. Ratiu [1998c] The
Euler-Poincar\'{e} equations in geophysical fluid dynamics, in
{\it Proceedings of the Isaac Newton Institute Programme on the
Mathematics of Atmospheric and Ocean Dynamics}, Cambridge
University Press (to appear).

\item Holm, D.D., J.E. Marsden, T.S. Ratiu, and A. Weinstein
[1985] Nonlinear stability of fluid and plasma equilibria,
{\it Phys. Rep.\/} {\bf 123}, 1--116.

\item Kato, T. [1984] Remarks on the zero viscosity limit for
nonstationary flows with boundary, {\it Seminar on Partial Differential
Equations}, S.S. Chern, Ed., Springer-Verlag, 85--98.

\item Klein, M.  [1970] {\it Paul Ehrenfest.\/}
North-Holland Pub. Co.

\item Kouranbaeva, S. [1999] The Camassa-Holm equation as a
geodesic flow on the diffeomorphism group, {\it J. Math. Phys.} {\bf 40},
857--868.

\item Markovitz, H., and Coleman, B.D. [1964]  Incompressible
second-order fluids, {\it Adv. Appl. Mech.}  {\bf 8},  69--101.

\item Marsden, J.E. [1973] On product formulas for nonlinear
semigroups. {\it J. of Funct. An.} {\bf 13}, 51--72.

\item Marsden, J.E. [1973]
A formula for the solution of the Navier-Stokes
equation based on a method of Chorin.
{\it Bull. AMS.} {\bf 80}, 154--158.

\item Marsden, J.E. [1992],
{\it Lectures on Mechanics\/} London Mathematical
Society Lecture note series, {\bf 174}, Cambridge University Press.

\item Marsden, J.E., D.G. Ebin, and A. Fischer [1972]
Diffeomorphism groups, hydrodynamics and relativity.
{\it Proceedings 13th Biennial Seminar on Canadian
Mathematics Congress\/}, 135--279.

\item Marsden, J.E. and T.J.R. Hughes [1983]
{\it Mathematical Foundations of Elasticity.\/}
Prentice Hall, reprinted by Dover Publications, N.Y., 1994.

\item Marsden, J.E. and T.S. Ratiu [1999]  {\it Introduction to
Mechanics and Symmetry.} Texts in Applied Mathematics {\bf 17},
1994, Second Edition (1999), Springer-Verlag.

\item Marsden, J.E., T.S. Ratiu and S. Shkoller [1999a]
The geometry and analysis of the averaged Euler equations
and a new diffeomorphism group, {\it Geom. and Funct. Analysis} (to
appear).

\item Marsden, Ratiu and Shkoller [1999b] On the averaged Euler
equations (in preparation).

\item Misiolek, G. [1993] Stability of flows of ideal fluids
and the geometry of the group of diffeomorphisms.  {\it Indiana
Univ. Math. J.\/} {\bf 42}, 215-235.

\item Misiolek, G. [1996] Conjugate points in ${\cal D}(T^2)$.
{\it Proc. Amer. Math. Soc.} {\bf 124} 977--982.

\item Misiolek, G. [1998] A shallow water equation as a
geodesic flow on the Bott-Virasoro group. {\it J. Geom. Phys.},
{\bf 24}, 203--208.

\item Moser, J. [1966] A rapidly convergent iteration method
and nonlinear partial differential equations, I {\it Ann. Sc.
Norm. Sup. Pisa\/} {\bf 20}, 265--315.

\item Noll, W., and C. Truesdell [1965] {\it The nonlinear
field theories of Mechanics}, Springer-Verlag, Berlin.

\item Palais, R.S. [1968]
{\it Foundations of Global Non-Linear Analysis.\/}
Benjamin/Cummins Publishing Co., Reading, MA.

\item Poincar\'{e}, H. [1890] {\it Th\'eorie des tourbillons},
Reprinted by \'Editions Jacques Gabay, Paris.

\item Poincar\'{e}, H. [1892] Les formes d'\'{e}quilibre d'une
masse fluide en rotation,  {\it Revue G\'{e}n\'{e}rale des
Sciences\/} {\bf 3}, 809--815.

\item Poincar\'{e}, H. [1901a] Sur la stabilit\'{e} de
l'\'{e}quilibre  des figures piriformes affect\'{e}es par une masse
fluide en rotation, {\it Philosophical Transactions A\/} {\bf 198},
333--373.

\item Poincar\'{e}, H. [1901b] Sur une forme nouvelle des
\'{e}quations de la m\'{e}chanique,  {\it C.R. Acad. Sci.\/} {\bf
132}, 369--371.

\item Poincar\'{e}, H. [1910] Sur la precession des corps
deformables. {\it Bull Astron\/} {\bf 27}, 321--356.

\item Rivlin, R.S., and J.L. Ericksen, [1955] Stress-deformation
relations for isotropic materials, {\it J. Rat. Mech. Anal.} {\bf
4},  323--425.

\item Rajagopal, K.R., and A.S. Gupta, [1981]  On a class
of exact solutions to the equations of motion of a second grade
fluid, {\it Inter. J. Engrg. Sci.} {\bf 19}, 1009--1014.

\item Shkoller, S. [1998] Geometry and curvature of diffeomorphism
groups with $H^1$ metric and mean hydrodynamics, {\it
J. Func. Anal.} {\bf  160}, 337--365.

\item Temam, R. [1988] {\it Infinite Dimensional Dynamical
Systems in Mechanics and  Physics}, Springer Verlag, New York, 1988.

\item Ting, T.W. [1963] Certain non-steady flows of second-
order fluids, {\it Arch. Rat. Mech. Anal.} {\bf 14}, 1-26.

\item Wang, C.Y. [1991] Exact solutions of the steady-state
Navier-Stokes equations, {\it Ann. Rev. Fluid Mech.} {\bf 23}, 159-177.

\end{description}

\end{document}